\documentclass[onecolumn,12pt,preprintnumbers,aps]{revtex4}
\begin{document}
\title{Lattice disorder and Ferromagnetism in La$_{0.67}$Ca$_{0.33}$MnO$_{3}$ nanoparticle}
\author{$^1$R.N. Bhowmik\footnote{E-mail address for correspondence:\\ rnbhowmik.phy@pondiuni.edu.in (RNB)\\ 
r.ranganathan@saha.ac.in (RR)}, $^2$Asok Poddar, $^2$R. Ranganathan, and $^2$Chandan Majumdar}
\affiliation{$^1$Department of Physics, Pondicherry University, R.V. Nagar, Kalapet, Pondicherry-605014, India\\
$^2$Experimental Condensed Matter Physics Division, Saha Institute of Nuclear Physics, 1/AF Bidhannagar, Kolkata-700064, India}

\begin{abstract}
We study the ferromagnetism of La$_{0.67}$Ca$_{0.33}$MnO$_{3}$ in bulk polycrystalline, nanocrystalline and amorphous phase. The structural change from crystalline phase to 
amorphous phase exhibited a systematic decrease of T$_C$(paramagnetic to ferromagnetic transition temperature) and spontaneous magnetization (M$_S$). The experimental results suggested few more features, e.g., appearance of large magnetic irreversibility in the temperature dependence of magnetization, lack of magnetic saturation at high magnetic field, blocking of magnetization below T$_B$, and enhancement of coercivity. In addition, the magnetic phase transition near to T$_C$ has changed from first order character in bulk sample to second order character in nanocrystalline and amorphous samples. We understand the observed magnetic features as the effects of decreasing particle size and increasing magnetic (spin-lattice) disorder. We noted that magnetic dynamics of amorphous samples is distinctly different from the nanocrystalline samples. The ferromagnetism of amorphous samples are comparable with the properties of reported amorphous ferromagnetic nanoparticles. We also demonstrate the effect of disorder shell in controlling the dynamics of ferromagnetic cores.  

\end{abstract}
\maketitle

\section{Introduction}
LaMnO$_3$ and CaMnO$_3$ are two typical antiferromagnets of perovskite structure with T$_N$ $\sim$ 140 K and 130 K, respectively \cite{Wolton}. The mixed compounds La$_{1-x}$Ca$_x$MnO$_3$ [(La$^{3+}_{1-x}$Ca$^{2+}_{x}$)(Mn$^{3+}_{1-x}$Mn$^{4+}_{x}$)O$^{2-}_3$] have shown a rich magnetic phase diagram both in hole doped (x $<$ 0.5) and electron doped (x $>$ 0.5) regions 
\cite{Wolton,Martin,Ramirez}. The hole doped and electron doped regions are dominated by Mn$^{3+}$(3d$^4$) and Mn$^{4+}$(3d$^3$) ions, respectively. The other important aspect of these manganites is to understand the colossal magnetoresistance (CMR) below room temperature \cite{CNR}. Although CMR properties have been discussed in terms of double exchange (ferromagnetic) interactions between Mn$^{3+}$-O-Mn$^{4+}$ ions, but double exchange (DE) interactions alone can not explain the complete mechanism of CMR effect. Recently, many magnetic (spin-lattice) disorder effects, e.g., Jahn-Teller distortion (due to the presence of Mn$^{3+}$), and electron-phonon coupling (due to the lattice strain and deformation of Mn$^{3+}$-O-Mn$^{4+}$ bonds), were considered for CMR manganites \cite{Font,Sun}. Most of the literature works have attempted to study the roles of magnetic disorder by cation (La or Mn) substitution \cite{Sun,Liu,Yusuf}. The dilution of magnetic cations (Mn) has shown the weakening of ferromagnetic DE interactions. Finally, a competition between ferromagnetic DE interactions (Mn$^{3+}$-O-Mn$^{4+}$) and  antiferromagnetic (Mn$^{3+}$-O-Mn$^{3+}$ and Mn$^{4+}$-O-Mn$^{4+}$) superexchange interactions may result in the freezing of magnetic domains or phase separation phenomena in manganites \cite{Yusuf,Dagotto}.\\
The present compound La$_{0.67}$Ca$_{0.33}$MnO$_3$ (LCMN) is a ferromagnet with T$_C$ reported over a range 250 K to 285 K \cite{Yusuf,Dagotto,Lynn,Rett}. The understanding of magnetic phase transition near T$_C$ of LCMN, whether first order or second order, is an issue of recent interest \cite{Robler,Kim,Mira,Heffner,Novak}. The study of ferromagnetic ordering of LCMN about T$_C$ could be relevant to get insight of the CMR behaviour in many magnetic oxides. La$_{0.67}$Ca$_{0.33}$MnO$_3$ is a Mn$^{3+}$ (3d$^4$) rich compound, where crystal field splits the five-fold degenerate 3d$^4$ levels into t$_{2g}^{3}$ and e$_{g}^{1}$ levels. The effective ferromagnetic coupling (J$_H$) between e$_g$ electrons (spin= 1/2) and t$_{2g}$ electrons (spin= 3/2) of on-site (Mn) is much more stronger than the hopping interaction (t$^{0}_{ij}$) of the e$_g$ electrons between holes of neighbouring 
Mn$^{4+}$ (3d$^3$) at sites, i and j \cite{Tokura}. The Anderson-Hasegawa relation 
\cite{Anderson} (t$_{ij}$ = t$^{0}_{ij}$cos($\theta_{ij}$/2)) suggests that the effective hopping interaction (t$_{ij}$) depends on the relative angle $\theta_{ij}$ between the spins of neighbouring sites.\\ 
In our opinion, the effect of $\theta_{ij}$ on the effective exchange interactions can be best studied by reducing the particle size of manganites into nanometer scale, where shell (surface) spins of the nanoparticle are expected to be more disordered in comparison with the core spins. Recently, spin polarized tunneling between two grains (cores) via grain boundaries (shells) \cite{Hwang,Lee} has been proposed to explain the large magnitude of low field magnetoresistance (MR) in manganite nanomaterials. A proper knowledge of magnetic interactions between core-shell spins would be useful to realize the inter-grain tunneling of polarized spins, as well as the effect of disorder on double exchange ferromagnetism. 
A few reports are available to demonstrate the underlying physics of CMR manganites by reducing the particle size in the nanometer scale \cite{Lopez,Mahesh,Sanchez,TKN,Lynn}.
A better understanding of magnetic disorder effects, related to the grain and grain boundaries, can be achieved by comparing the properties of a CMR manganite in different structural (i.e., bulk polycrystalline, nanocrystalline and amorphous) phase 
\cite{Biasi,Proz}. It will be interesting to study, to what extent, the properties of disordered magnets, such as spin glass, cluster glass, superparamagnetism and short-ranged magnetic ordering above T$_C$, are affected by the structural disorder in the amorphous phase of a ferromagnetic manganite. In the present work, we have highlighted such magnetic aspects in 
La$_{0.67}$Ca$_{0.33}$MnO$_3$ as a function of particle size, both in crystalline and  amorphous phase.
\section{Experimental} 
The stoichiometric amounts of high purity La$_2$O$_3$, CaCO$_3$, MnO$_2$ oxides have been  mixed to obtain the bulk La$_{0.67}$Ca$_{0.33}$MnO$_3$. The mixture has been ground for 2 hours and pelletized. The pellets have been sintered at 1100$^0$C for 12 hours, at 1250$^0$C for 20 hours and at 1380$^0$C for 20 hours with several intermediate grindings by cooling 
the sample to room temperature. Finally, the sample is cooled to room temperature at 2-3$^0$C/min. The X ray diffraction (XRD) spectrum at 300 K, using Philips PW1710 diffractometer with Cu K$_{\alpha}$ radiation, confirmed the formation of bulk La$_{0.67}$Ca$_{0.33}$MnO$_3$ (LCMN) sample. The XRD data have been recorded in the 2$\theta$ range 10-90$^0$ with step size 0.01$^0$. Fritsch Planetary Mono Mill "Pulverisette 6" has been used for mechanical milling of powdered LCMN sample. The milling has been carried out in Argon atmosphere in an 80 ml agate vial with 10 mm agate balls. The ball to sample mass ratio was maintained to 7:1. The agate bowl and balls have been selected to avoid the magnetic contamination to the material, if happens at al during milling process. The samples with specific milling hours (X) are denoted as mhX. The structural phase evolution during milling process has been checked from the XRD spectrum of each sample. Crystal structure of the samples have been analyzed by standard full profile fitting method using FULLPROF Program. Particle size of the samples have been calculated using Transmission Electron Microscope (TEM) (model: Tecnai S-twin). The surface morphology of samples (pellet form) have been studied by scanning electron microscope (SEM) (model: Hitachi S-3400N). Elemental composition of the samples have been determined from the energy dispersive analysis of X-ray (EDAX) spectrum (Thermo electron corporation Instrument), attached with SEM.\\
ac susceptibility and dc magnetization of the samples have been measured in the 
temperature range 5 K to 340 K using SQUID magnetometer (MPMS-Quantum Design, USA). 
The real ($\chi^{\prime}$) and imaginary $\chi^{\prime\prime}$) part of ac susceptibility have been recorded in the frequency range 1 Hz-1.5 kHz at ac field amplitude h$_{rms}$ $\approx$ 
1 Oe. The dc magnetization (M) measurement as a function of temperature (T) has been performed with conventional zero field cooled (ZFC) and field cooled (FC) modes. In ZFC mode, the sample has been zero field cooled from 340 K down to 5 K. Then, M(T) data are recorded in the presence of measurement field while temperature increases from 5 K. In FC mode, the sample has been cooled from 340 K to 5 K in the presence of magnetic field and the cooling field is maintained at the same value during the M(T) measurement from 5 K to 340 K. The field dependence of magnetization (M(H)) has been carried out by cooling the sample at the measurement temperature under ZFC mode. 

\section{RESULTS}
\subsection{Structural phase}
Fig. 1a shows the room temperature XRD spectra of selected samples. The XRD spectra indicated that crystalline nature of the material decreases significantly for the milling time more than 61 hours and amorphous phase dominates in the spectrum for milling time more than 98 hours. The milled samples upto 98 hours milling time do not show any additional peaks in comparison with the spectrum of bulk sample. This suggests that both bulk and milled samples (upto mh98) are in similar crystallographic phase and found to be matching with orthorhombic structure with Pnma space group. The lattice parameters are shown in Table I. The lattice parameters ($\it{a}$, $\it{b}$ and $\it{c}$) of bulk sample is within the range of reported values (a$\sim$5.4467$\AA$, b$\sim$7.6914$\AA$, c$\sim$5.4616$\AA$ \cite{Sagdeo}; a$\sim$5.4289$\AA$, b$\sim$7.8194$\AA$, c$\sim$5.4557$\AA$ \cite{Liou}). We have seen the decrease of both $\it{a}$ and $\it{c}$ with milling time, whereas $\it{b}$ increases. Over all, cell volume increases with the increase of milling time. The XRD line broadening in milled samples can be attributed to an effect of either increasing amorphous phase or decreasing particle size in the material.  To get a quantitative estimation of retaining crystalline phase in the milled samples, the intensity of (200) peak line of all the samples is divided by (200) peak intensity of bulk sample (Fig. 1b). The relative peak intensity (shown in Table I) is significantly reduced for milling time more than 98 hours and approaching to zero value at higher milling time (e.g., $\sim$ 3\% for mh200 sample). The structural change can be followed by denoting the bulk sample as polycrystalline (peak intensity 100\%), the sample with relative peak intensity between 90-10 \% as nanocrystalline, the sample with relative peak intensity $\leq$ 5 \% as amorphous, and the sample with relative peak intensity between 9-6 \% as nanocrystalline phase coexisting with amorphous phase. 
On the other hand, average size of the material (shown in Table I) from TEM micrographs is not monotonically decreasing with the increase of milling time (size $\sim$ 65 nm, 12 nm and 90 nm for mh25, mh61, and mh200 samples, respectively). The results suggested the appearance of grain growth kinetics at milling time $\geq$ 98 hours, whereas the appearance of a 
large fraction of amorphous phase after milling time $\geq$ 98 hours is evidenced from the broad hump in XRD spectrum. The SEM pictures (Fig. 2a-c) suggest a relatively smooth surface of bulk sample in comparison with  milled samples. The change in surface morphology is understood from the breaking 
of $\mu$m size (1$\mu$m to 10$\mu$m) particle size of bulk sample into the nanometer range for mechanical milled samples. This is confirmed from SEM pictures, as well as TEM pictures. The estimated size of particles from SEM pictures is larger due to agglomeration of particles. 
The elemental composition and impurity effect during mechanical milling have been checked from EDAX spectrum. Fig. 2d-f shows the spectrum of selected samples. The spectrum has been recorded at 10 different points for each sample over a zone of 125$\mu$m x 125$\mu$m. The spectrum indicated the presence of La, Ca, Mn and O as main elements in the samples. 
The spatial homogeneity of elements is also seen from the elemental mapping (data not shown). 
The elemental composition (La$_{0.67\pm0.01}$Ca$_{0.33\pm0.01}$MnO$_{3-\delta}$) of the 
samples is close to the expected compound La$_{0.67}$Ca$_{0.33}$MnO$_3$. The analysis indicated slight oxygen deficiency in our samples with $\delta$ $\sim$ 0.1 for mh200 sample. We do not find any significant amount of impurity during mechanical milling for mh61 sample. However, a small amount (less than 3 atomic \%) of Si impurity is detected in mh200 sample. 

\subsection{ac susceptibility}
The real ($\chi^{\prime}$) and imaginary ($\chi^{\prime\prime}$) parts at 10 Hz for selected samples are shown in Fig. 3. The $\chi^{\prime}$ of bulk sample increases sharply at about 
T$_m$ $\sim$ 270 K, followed by a slow decrease on lowering the temperature. The $\chi^{\prime\prime}$ shows two peaks at T$_{1}$ (= T$_m$) $\sim$ 270 K 
and at T$_{2}$ $\sim$ 40 K, respectively. It may be noted that there is no significant change in $\chi^{\prime}$ near 40 K. The sharp increase in $\chi^{\prime}$ and associated peak 
in $\chi^{\prime\prime}$ at about 270 K for bulk sample changes into a broad maximum 
for milled samples. Both $\chi^{\prime}$ and $\chi^{\prime\prime}$ decreases below 
the temperature of ac susceptibility maximum (T$_m$ for $\chi^{\prime}$ and T$_{1}$ for $\chi^{\prime\prime}$). The $\chi^{\prime\prime}$ of milled samples does not show 
any low temperature peak, as seen in bulk sample at 40 K. The important change, we noted here, is the shift of both T$_m$ (position of $\chi^{\prime}$ maximum) and T$_{1}$ (position of $\chi^{\prime\prime}$ maximum) to lower temperatures with milling time. 
The shift is much clear in T$_{1}$ as compared to T$_m$. A gradual transformation in  
the magnetic dynamics of milled samples is indicated from the fact that 
T$_{1}$ (position of $\chi^{\prime\prime}$ maximum) is higher than T$_{m}$ (position of $\chi^{\prime}$ maximum) for milling time upto 98 hours (Fig. 3 (a-c)) and opposite is true (i.e., T$_{1}$ $<$ T$_{m}$) for milling time more than 98 hours (Fig. 3 (d-e)).  
In order to understand the magnetic dynamics (spin ordering) for the samples with milling time less and more than 98 hours, we have studied the ac susceptibility at different frequencies 
($\nu$ = 1 Hz to 1.5 kHz) for mh61 (nanocrystalline) and mh146 (amorphous) samples (Fig. 4). The magnitude of $\chi^{\prime}$ and $\chi^{\prime\prime}$ for mh61 sample (Fig. 4a) have shown certain changes with the frequency variation. There is a branching of ac susceptibility data in the temperature range 40 K to 200 K. The magnitude of $\chi^{\prime}$ decreases with the increase of frequency, unlike the increase of $\chi^{\prime\prime}$ magnitude. Such frequency dependence of ac susceptibility indicates the presence of finite magnetic disorder  in the sample. The absence of a clear frequency shift either at T$_{m}$ $\sim$ 165 K or at 
T$_{1}$ $\sim$ 200 K suggests that magnetic disorder in mh61 sample is not sufficient for 
exhibiting a typical spin glass freezing or superparamagnetic blocking. On the other hand, 
ac susceptibility ($\chi^{\prime}$, $\chi^{\prime\prime}$) maximum of mh146 (amorphous) 
sample shows a clear shift with frequency (Fig. 4b). The determination of T$_{m}$($\nu$) 
shift from $\chi^{\prime}$ may not be much accurate due to more broadening of maximum. Hence, we have analyzed the frequency shift of $\chi^{\prime\prime}$ maximum at T$_{1}$($\nu$) (T$_1$ $\sim$ 102 K, 108 K, 114 K, 119.5 K, 121.5 K and 123 K for 1 Hz, 99 Hz, 575 Hz, 997 Hz and 1465 Hz, respectively). The temperature shift per decade of frequency ($\Delta$T$_{1}$($\nu$)/T$_{1}$(1 Hz)$\Delta$ln($\nu$)) is $\sim$ 0.028. The fit of T$_{1}$($\nu$) according to Vogel-Fulcher law: $\nu$ = $\nu_0$exp(-E$_a$/k$_B$(T$_{1}$-T$_0$)) is shown in the inset of Fig. 4b. The fit parameters are $\nu_0$ (characteristic spin flip frequency) = 8.4$\times$10$^{14}$ Hz, E$_a$ (activation energy)= 0.23 ev and T$_0$ (interaction parameter) = 25 K. 
The values of fit parameters are not consistent with a typical superparamagnetic blocking of magnetic particles (where T$_0$ = 0 K, $\nu_0$ $\sim$ 10$^{10}$-10$^{12}$ Hz, temperature shift per decade of frequency $\sim$ 0.1) about the ac susceptibility maximum of mh146 sample. The non-zero value of T$_0$ indicates a sufficient amount of interactions among the magnetic particles. The obtained value of temperature shift per decade of frequency is also larger than the typical value 0.001 for classical spin-glasses like CuMn. However, the fit parameters of mh146 sample are comparable with other amorphous nanomagnets ($\nu_0$ $\sim$10$^{8}$-10$^{12}$ Hz, T$_0$ $\sim$ 25 K-35 K) \cite{Mukum,Toro}, which have shown a collective freezing of interacting magnetic clusters at lower temperatures. 

\subsection{dc magnetization}
\subsubsection{Temperature dependence of dc magnetization}
The temperature dependence of ZFC and FC magnetization at 100 Oe are plotted in 
Fig. 5. Some of the features of zero field cooled magnetization (MZFC) are similar to the 
observations in the real part ($\chi^\prime$) of ac susceptibility. 
For example, a sharp increase of magnetization about 275 K, followed by a slow decrease of MZFC for bulk sample, and the appearance of a broad maximum in MZFC, followed by a sharp decrease below 50 K for milled samples were observed in ($\chi^\prime$)(T) data (Fig. 3).
There is no difference between MZFC and field cooled magnetization (MFC) above 270 K. 
MFC separates out from MZFC below 270 K and the magnitude of MFC increases down to 10 K. 
The irreversibility temperature T$_{irr}$ decreases with the increase of milling time 
(e.g., T$_{irr}$ $\sim$ 265 K, 259 K, 245 K, 225 K, 210 K, and 190 K for bulk, mh25, mh61, mh98, mh146 and mh200 samples, respectively). The irreversibility effect at 100 Oe is qualitatively demonstrated (inset of Fig. 5) from the temperature dependence of normalized 
thermoremanent magnetization NTRM (=[MFC(T)-MZFC(T)]/TRM(10K)). It is interesting to note that TRM (10 K) of nanocrystalline samples (4.87 emu/g, 3.55 emu/g, 2.31 emu/g for mh25, mh61, mh98, respectively) are larger than bulk sample (1.84 emu/g). On the other hand, TRM (10 K) 
of amorphous samples (1.32 emu/g for mh146 and 0.70 emu/g for mh200) not only lower than bulk sample, but also decreases with the increase of amorphous character in the material. The decrease of normalized TRM with the increase of temperature may be a general feature of magnetic materials, but a different kind of magnetic dynamics is marked in the change of shape of NTRM(T) curves from down curvature for milling time $\geq$ 98 hours to up curvature for milling time $\leq$ 61 hours. \\ 
We understand the nature of magnetic exchange interactions from the analysis of temperature dependence of inverse of dc susceptibility ($\chi^{-1}$ = H/M) curves at H = 100 Oe (Fig. 6). The high temperature (T $\geq$ 300 K) data of bulk sample are fitted with simple Curie-Weiss law:
\begin{equation}
 \chi = C/(T-\theta_w)
\end{equation} 
In the expression, C ($\sim$ 0.0176) is Curie constant and $\theta_w$ ($\sim$ 275 K) is paramagnetic Curie temperature. The effective magnetic moment ($\mu$ $\sim$ 5.42 $\mu_B$ per f.u.) is estimated from the Curie constant (C= N$\mu^2$/3k$_B$, N is the number of formula unit (f.u.)). The estimated value of $\mu$ is large compared to 4.56$\mu_B$ per Mn atom for the spin contribution alone (assuming 67\% Mn$^{3+}$ with S =2 and 33\% Mn$^{4+}$ with S = 3/2 and g$_S$ = 2, as suggested in Ref. \cite{Wolton}). Another aspect is that the non-linear (slightly up curvature) variation of $\chi^{-1}$ (T) data above 260 K for bulk sample can be fitted using 
\begin{equation}
\chi = C/(T-\theta_w)^{\gamma}
\end{equation}
 with $\gamma$ = 1.20 and $\theta_w$ $\sim$ 260 K. It is remarkable to note that $\chi^{-1}$ (T) plots of milled samples show different character in comparison with bulk sample. 
First, we fit the data (T$>$275 K) using simple Curie-Weiss law. The fit parameters of Curie-Weiss law ($\mu$, $\theta_w$) are shown in Table II. The hyperbolic shape of 
$\chi^{-1}$ (T) (with down curvature) below 300 K suggests that milled samples belong to the class of either ferrimagnet or double exchange ferromagnet \cite{Anderson}. 
The $\chi^{-1}$ (T) data in the temperature range 240 K-340 K are fitted with the following equation, generally applied for ferrimagnet 
\cite{ferri}.
\begin{equation} 
1/\chi = (T-\theta_{1})/C_{eff} - \xi/(T-\theta_{2})
\end{equation}
In our fitting method, we freely allow the parameters ($\theta_{1}$, C$_{eff}$ ($\mu_{eff}$), $\xi$ and $\theta_{2}$) to take initial values. As soon as the fitted curve comes close to the experimental curve, we start to restrict the parameters one by one, except $\xi$, to obtain the best fit curve. The main panel of Fig. 6 suggests a good quality fit according to equation (3). The fit parameters are shown in Table II. A comparative fits of equation (1) and (3) for mh61 sample suggests that equation (1) may be well valid at higher temperature, but equation (3) is more appropriate to describe the magnetic behaviour over a wide temperature range. The interesting point is that the change of effective moment with the milling time is identical, as obtained from simple Curie-Weiss law (equation 1) and from equation (3), except the magnitude of $\mu_{eff}$ is larger than $\mu$. In both cases, the effective moment increases with milling time upto 61 hours (nanocrystalline samples) and for milling time $\geq$ 98 hours the effective moment, again, decreases and attaining a constant magnitude for the amorphous (mh146 and mh200) samples. The relatively large effective magnetic moment from equation (3) may be involved to the accuracy of fit parameters in the limited temperature range or the clustering of moments in the paramagnetic state \cite{Ma,Tera}. This limits the use of $\mu_{eff}$ for quantitative discussion. However, $\mu$ can be used for the qualitative understanding of paramagnetic moment with milling time.

\subsubsection{Field dependence of dc magnetization}
The field dependence of magnetization of the samples are studied at different temperatures. The magnetization is measured at 10 K with field range $\pm$ 70 kOe, but M(H) data are shown
 within H= $\pm$ 3 kOe for the clarity of Fig. 7a. We have noted some typical features of coercive field H$_C$ (where magnetization reverses its sign), irreversible field H$_{irr}$ (where the loop open below this specified field) and remanent magnetization M$_R$ (which represents the residual magnetism after making the field to zero value either from +70 kOe or -70 kOe) and the energy product (proportional to the hysteresis loss or loop area), defined by H$_C$xM$_R$. Both H$_C$ and H$_{irr}$ (in Fig. 7b) showed a rapid increase in the initial stage of milling time (nanocrystalline samples) and then, approaching to a constant value after 98 hours (amorphous samples). The increase of H$_C$ is related to the increasing anisotropy energy of the material, whereas the increase of H$_{irr}$ is related to the increasing magnetic disorder in the material. The initial increase of M$_R$ for mh25 sampple is followed by a gradual decrease with milling time $\geq$ 61 hours. However, energy product (H$_C$xM$_R$) for the nanocrystalline samples is large compared to bulk sample, but decreases for the samples in amorphous phase. Similar variation of energy product has been found in magnetic nanocomposites, consisting of soft (core) and hard (shell) components \cite{Coey}. 
Inspite of the same order of energy product for both bulk and amorphous (mh200) samples, the experimental result indicated that coercivity of soft ferromagnet ($\sim$ 30 Oe for bulk) can be enhanced ($\sim$ 540 Oe for mh200) by making the material in amorphous phase.\\
 Fig. 8 represents the M(H) curve at different temperatures (10 K to 300 K) for selected samples. A typical soft ferromagnetic character of bulk sample is confirmed from the 
rapid increase of magnetization within 5 kOe field and a tendency to attain magnetic saturation at higher fields. At the same time, a field induced magnetic behaviour is indicated from the small M-H loop (Fig. 8(b)) at about 30 kOe and 50 kOe for 10 K and 30 K, respectively, while the field was reversing back from +70 kOe. There is no field induced transition at T $\geq$ 50 K. A close look at the M(H) data suggested that magnetization of 
bulk sample is not completely saturated even at 70 kOe and the signature of magnetic non-saturation is more prominent at T $\geq$ 260 K. The non-saturation character of high field magnetization clearly dominates in the material when the lattice structure changes from nanocrystalline phase to amorphous phase (Fig. 8: (c) mh98, (d) mh200).\\
In order to determine the spontaneous magnetization (M$_S$) and paramagnetic to ferromagnetic transition temperature (T$_C$), we have analyzed M-H data using Arrot plot 
(M$^2$ vs. H/M) \cite{Mira}. The normalized Arrot plot (M(H)$^2$/M(70 kOe)$^2$ vs. H/M ) at 10 K is shown in Fig. 9a. The linear extrapolation of high field data intercepts on the positive 
M$^2$ axis for bulk, mh25 and mh61 samples, and the determination of M$_S$ is easy. On the other hand, a non-linear curve intercepts on the positive M$^2$ axis for the samples 
with higher milling time. This is the signature of increasing disorder effect in the material 
\cite{rnbMnCr2O4}. We obtain the spontaneous magnetization for these samples from 
the polynomial fit of high field data. The polynomial fit is shown in Fig. 9a for mh98 and mh200 samples. The extrapolation of M$_S$ (T) curve to M$_S$ = 0 value determines T$_C$ of 
the sample, whereas the extrapolation to T = 0 K gives the spontaneous magnetization M$_S$(0). The values of T$_C$ and M$_S$ (0) are shown in Table II. Our experimental results show 
that both T$_C$ and M$_S$(0)decreases with milling time, which indicates the loss of 
long range parameters in La$_{0.67}$Ca$_{0.33}$MnO$_{3}$ ferromagnet. We understand the order of magnetic phase transition near T$_C$ of the samples from H/M vs. M$^2$ isotherms, 
which has been recognised as an useful method \cite{Kim,Mira}. The extrapolated 
isotherms intercept the M$^2$ axis with positive slope (shown by dotted line in Fig. 9b) at 
T$\leq$ 260 K for bulk sample. In addition to the positive slope for high field isotherms, a negative slope also intercepts the M$^2$ axis at lower fields in the temperature range 270 K to 280 K. No positive slope intercepts the M$^2$ axis at 285 K, which is above the T$_C$ 
$\sim$ 281 K of bulk sample. On the other hand, only a positive slope intercepts the M$^2$ axis up to T$_C$ for all milled samples (Fig. 9c-d), irrespective of nanocrystalline or amorphous structure. It is established \cite{Kim,Mira,Good} that negative and positive slope gives the signature of first order and second order character, respectively. This suggests a drastic change in the order of paramagnetic to ferromagnetic phase transition in milled samples. Fig. 10 showed the lowering of reduced magnetization curves(M$_S$(T)/M$_S$(0) vs. T/T$_C$) for milled samples in comparison with bulk sample. The reduced curves qualitatively  suggested more disorder in milled samples and indicated the increasing fluctuations about 
mean ferromagnetic exchange interactions \cite{Poon}, when lattice structure of the bulk ferromagnet changes into nanocrystalline and amorphous phases. The Rhodes-Wohlfarth (P$_C$/P$_S$ vs. T$_C$) plot \cite{Hardy} plot is shown in the inset of Fig 10, where P$_S$ = M$_S$(0) per f.u. and effective magnetic moment per f.u. $\mu_{eff}^{2}$ = P$_C$(P$_C$+2) gives P$_C$. 
The plot shows that P$_C$/P$_S$ for all the samples is larger than the typical value 1 for pure localized magnetism. The increase of P$_C$/P$_S$ with the decrease of T$_C$ (P$_C$/P$_S$= 1.25, 46.29 and T$_C$= 281 K, 212 K for the bulk and mh200 samples, respectively) suggests the probability of increasing itinerant character in the ferromagnetic properties of milled samples in comparison with localized magnetism of bulk sample \cite{Riva2}.   

\section{DISCUSSION}
We have employed mechanical milling to transform bulk La$_{0.67}$Sr$_{0.33}$MnO$_{3}$ into different lattice structures. We preferred discontinuous milling of the material. The milling procedure was stopped in every 6 hours interval for proper mixing of the milled powder and to minimize the agglomeration effect of particles. We are able to reduce the particle size down to 12 nm for milling time 61 hours and grain growth kinetics appeared at higher milling time. In contrast, there is a systematic change of lattice structure with milling time from bulk polycrystalline to nanocrystalline and amorphous phase. It is interesting to note that  particle size of mh25 (65 nm) and mh146 samples (60 nm) are nearly same order, but mh25 sample is more crystalline (87\%) than mh146 sample (5\%) and their magnetic properties are  drastically different from each other. This indicates that local heating during mechanical milling upto 200 hours is not sufficient to exhibit the milling induced crystallization effect \cite{Kown}. A small amount of Si contamination (less than 3 atomic \%) from the agate bowl and balls has been observed at higher milling time (200 hours). Considering the non-magnetic character of Si atoms, the effect of such small amount of impurity can be treated as some additional disorder in the lattice structure whose magnetic effect is negligible. In fact no significant amount of contamination is found for mh61 sample, but magnetic properties are significantly changed with respect to bulk sample. Hence, we discuss the ferromagnetism of La$_{0.67}$Ca$_{0.33}$MnO$_{3}$ in terms of particle size reduction and lattice disorder.\\
First, we understand the magnetic ordering of bulk La$_{0.67}$Ca$_{0.33}$MnO$_{3}$. The T$_C$  is, generally, determined from the ac susceptibility peak temperature or the temperature where dc magnetization sharply increases or the temperature where spontaneous magnetization becomes zero and all these temperatures are nearly same for a long ranged ferromagnet. 
In the present work, ac susceptibility ($\chi^{\prime\prime}$) shows a sharp peak at $\sim$ 270 K, the dc magnetization (at 100 Oe) sharply increases at about 275 K, and M$_S$(T) data indicated a non-zero value of spontaneous magnetization below 281 K. These results clearly suggest that  bulk sample is not in pure ferromagnetic phase below T$_C$ $\sim$ 281 K (determined from M$_S$(T) data). This is also confirmed from other experimental facts. For example, the fit exponent $\gamma$ = 1.22 from equation (2) is larger than the mean field theory predicted value ($\sim$ 1) for second order magnetic phase transition. However, the obtained value of $\gamma$ falls in the range 1 to 1.36, which has been reported for first order magnetic transition \cite{Robler,Kim}. The next experimental evidence is the nature of magnetic phase transition, i.e., second order character at T $\leq$ 260 K (positive slope in H/M vs M$^2$ plot) gradually changes into the first order character at T$_C$ $\sim$ 281 K (negative slope in H/M vs M$^2$ plot). The mixture of first order and second order character in between 270 K to 280 K represents a diffused magnetic state below T$_C$, just like liquid-gas mixture below critical temperature of the thermodynamic phase diagram. Our observations are consistent with literature reports \cite{Heffner,Burgy,Dagotto,Lyn,Novak}, which suggested that coexisting clusters in the ferromagnetic ground state are playing a major role to show the first order character of magnetic phase transition. The existence of magnetic clusters in the ferromagnetic matrix of our bulk sample is understood from the appearance of magnetic irreversibility between MZFC and MFC, and non-saturated character of magnetization even at 70 kOe. The number of short ranged interacting clusters increases when the system approaches to the paramagnetic state and reflected in a field induced magnetic state in the temperature range 270 K-280 K, as reported earlier 
\cite{Ma,Mira2,Riva}. Our magnetization data clearly indicated that most of these 
clusters are melting in the ferromagnetic matrix due to strong ferromagnetic interactions 
\cite{Zain,Mira2,Novak} when the temperature decreases below 270 K. The remaining fraction of clusters freezes into local magnetic states at lower temperature, which is evidenced from the $\chi^{\prime\prime}$ peak at $\sim$ 40 K. This is due to the high sensitivity of $\chi^{\prime\prime}$ (T) to energy loss related to rotation/freezing of magnetic clusters and the fact is confirmed from the field induced magnetic behaviour at T $\leq$ 40 K. The absence of such anomaly in $\chi^{\prime}$ suggests a few  numbers of freezing clusters in the bulk sample.\\ 
Now, we demonstrate the magnetism of mechanical milled samples in nanocrystalline and amorphous phase. We noted some significant features which indicated a gradual change in the ferromagnetic behavour of milled samples. For example, the low temperature $\chi^{\prime\prime}$ peak at 40 K, as seen for bulk sample, is not appeared for milled samples, except a shoulder around the same temperature for nanocrystalline samples. The effects of increasing magnetic disorder with milling time are realized from many observations: (1) increased magnetic irreversibility between ZFC and FC magnetization, (2) appearance of a round shape maximum in MZFC curves, (3) differenet character (down curvature) of $\chi^{-1}$ (T) data in the paramagnetic state, and (4) appearance of upturn in the Arrot plot.  
Here, we would like to focus on the fit of $\chi^{-1}$ (T) data, which seems to be more interesting for revealing the dynamics of milled samples. The paramagnetic Curie temperature ($\theta_w$), obtained from the fit of equation (1), is positive for all the milled samples, except the decreasing magnitude with the increase of milling time (e.g., $\theta_w$ $\sim$ 240 K and 30 K for mh25 and mh200 samples, respectively). The positive value of $\theta_w$ suggested that double exchange ferromagnetic interactions are, still, sfficiently strong \cite{Anderson} in the milled samples of La$_{0.67}$Ca$_{0.33}$MnO$_{3}$. On the other hand, equation (3) is applicable for ferrimagnetic materials, consisting of two magnetic sublattices of antiparallel directions \cite{ferri,rnbMnCr2O4}, where extrapolation of high temperature $\chi^{-1}$ (T) data is expected to intersect the temperature axis at negative value. Table II shows that the fit parameter $\theta_1$ is not 
negative for all the milled samples, but a systematic decrease is found as the milling time increases from 25 hours ($\theta_1$ $\sim$ 150 K) to 200 hours ($\theta_1$ $\sim$ -70 K). 
The decrease of $\theta_1$ suggests the reduction of ferromagnetic (FM) exchange interactions (or development of antiferromagnetic (AFM) exchange interactions) in nanocrystalline and amorphous samples. The spin glass like feature in amorphous (mh146) sample clearly proves the 
development of antiferromagnetic exchange interactions in the material, because spin glass like feature needs sufficient amount of both magnetic disorder and competition between FM/AFM interactions.\\
The validity of ferrimagnetic equation (3) in our milled samples can be examined by considering the core-shell spin structure of nanoparticles. The shell spin structure is magnetically disordered in comparison with the ferromagnetic ordered core spins. The shell spins may not be typical antiparallel with respect to core, but effective spin moment of shell is obviously low in comparison with ferromagnetic core \cite{Zhang}. This allows us to compare the magnetic contributions form shell and core of a nanoparticle with two unequal magnetic sublatticles of a ferrimagnet. Let us try to understand the possible disorder effects in the core-shell structure of nanoparticles. The increasing antiferromagnetic interactions in the material can be associated with the spin canting and breaking of double exchange ferromagnetic bonds (Mn$^{3+}$-O-Mn$^{4+}$) in the shell part of nanoparticle. The reduction of double exchange bonds may be affected due to slight oxygen deficiency in milled samples, as seen from EDAX data. The oxygen deficiency has the effect to decrease Mn$^{4+}$/Mn$^{3+}$ ratio, which results in the possible increase of antiferromagnetic (Mn$^{3+}$-O-Mn$^{3+}$) superexchange interactions \cite{Dorr}. The experimental results of Zhao et al. \cite{Zhao} suggested no alternation of T$_C$ in La$_{0.5}$Ca$_{0.5}$MnO$_{3-\delta}$ ($\delta$$\leq$0.17) sample with the variation of oxygen deficiency ($\delta$). However, we observe a drastic reduction of T$_C$ ($\sim$ 281 K for bulk, 212 K for mh200) in La$_{0.67}$Ca$_{0.33}$MnO$_{3}$ with increasing milling time. This means slight oxygen deficiency may have a minor role in disturbing the double exchange ferromagnetic interactions in the present sample, but its effect is not enough to explain all the observed magnetic change in nanocrystalline and amporphous samples. Another source of disorder in the material may be originated from the mechanical strain induced effect 
\cite{Ding}, confined in the shell of nanoparticle. A typical variation of coercivity in our samples (i.e., initial increase is followed by almost no change at higher milling time) suggests that strain induced disorder is saturated at higher milling time, whereas the magnetic behaviour of amorphous phase samples (milling time more than 98 hours) is found to be markedly different from the nanocrystalline samples (milling time more than 98 hours). Hence, mechanical strain induced disorder may not be the sufficient cause for increasing fraction of antiferromagnetic exchange interactions/disorder with the increase of milling time that is suggested from the fit of equation (3). Here, we would like to state that coercivity of the material may be saturated at higher milling time, but irreversible field H$_{irr}$ is continuously increasing with milling time. 
Cosidering irreversible field H$_{irr}$ as a good indicator of disorderness in magnetic nanoparticle, we suggest that imparted mechanical energy at higher milling time is being used to create mainly lattice disorder in the material. The fact is confirmed from the structural transformation of nanocrystalline phase into amorphous phase for milling time more than 98 hours. On the other hand, similar type variation of coercivity was observed in many materials \cite{Sort}, consisting of ferromagnetic particles surrounded by antiferromagnetic matrix. 
Coupling the core-shell structure of nanopaticles \cite{Zhang} and increasing lattice disorder in the material, we assume that a large number of ferromagnetic clusters (cores) are coexisting in the disordered matrix (contributed by shell spins and lattice disorder) of mechanical milled (nanocrystalline and amorphous) samples. This picture is different from the bulk sample where a few number of short-ranged interacting clusters coexist in the long ranged ferromagnetic matrix. We propose a schematic diagram in Fig. 11 to represent the change of core-spin configuration as a consequence of increasing disorder in the material. We have also compared the spin-lattice configuration of polycrystalline sample (order in both spin and lattice) (Fig. 11a) with nanocrystalline (shell spin disorder and small lattice order) (Fig. 11b) and amorphous (disorder in both spin and lattice) (Fig. 11c) samples. We suggest that   ferromagnetic coherent length is confined within the cores and ferromagnetic order is disturbed in the shell part. On the other hand, lattice disorder is spreading in both shell and core parts of the nanoparticles.\\
The above discussion suggests that the decrease of particle size in the nanometer range alone is not sufficient to controll the magnetic properties of La$_{0.67}$Ca$_{0.33}$MnO$_{3}$ in nanocrystalline and amorphous phase, but increasing lattice disorder also playing an important role in determining the magnetic parameters. The experimental results indicated that ferromagnetic (spin-spin) interactions are sufficiently strong even in the amorphous phase of the material. The low temperature decrease of magnetization (both in ac and dc susceptibility) occurs due to the blocking/freezing of ferromagnetic cores in the disordered matrix \cite{Dorr}. The is significant to note that blocking behaviour becomes prominent for higher milling time as an effect of increasing lattice disorder, but not due to an effect of decreasing particle size. The rapid increase of lattice disorder causes a sharp decrease of spin-lattice interactions with milling time and controlling a drastic reduction of T$_C$ ($\sim$ 281 K for bulk to 212 K for mh200) and spontaneous magnetization at 0K ($\sim$ 3.6 $\mu_B$ for bulk to 0.1 $\mu_B$ for mh200). To our knowledge it is not clear, why first order ferromagnetic phase transition of bulk LCMN sample transforms into second order character in nanomaterials. The existence of short-ranged interacting clusters in the ferromagnetic matrix may be playing a major role for first order phase transition in bulk LCMN sample. This is not true for milled samples, because first order character is not seen, although increasing number of magnetic clusters (cores) is realized from the magnetic measurements. The nature of magnetic phase transition may be very much sensitive to the disturbance of spin-lattice interactions of bulk sample \cite{Rett}. The origin of magnetic phase transformation in our milled samples is, most probably, related to the dilution of spin-lattice interactions and change in cluster(core)-matrix (shell) configuration. The Rhodes-Wohlfarth (P$_C$/P$_S$ vs. T$_C$) plot suggested that itinerant character of La$_{0.67}$Ca$_{0.33}$MnO$_{3}$ ferromagnet increases with milling time. Inspite of the non-monotonic variation of $\mu$ (hence P$_C$) with milling time, the ratio of P$_C$/P$_S$ continuously increases. This suggests that decrease of P$_S$, mainly from shell spins, played a dominant role in determining P$_C$/P$_S$ and increasing fraction of shell spins may be the probable source of itinerant spin moments. The experimental evidence of itinerant ferromagnetic character due to shell spins may be realized from the large low field magnetoresistance (MR) in manganite nanomaterials, as reported in Ref. \cite{Hwang,Lee}.
The itinerant character of shell spins is activated by the weakening of spin-lattice coupling \cite{Lsheng} and  excitation of spins to higher energy states as an effect of mechanical strain induced anisotropy. The other magnetic features of the amorphous samples, i.e., lack of magnetic saturation and magnetic irreversibility, are consistent with reported amorphous ferromagnetic nanoparticles \cite{Biasi,Proz}. 

\section{CONCLUSIONS}  
La$_{0.67}$Ca$_{0.33}$MnO$_{3}$ ferromagnet exhibited many interesting features in the nanocrystalline and amorphous phase, as an effect of increasing disorder in core-shell spin morphology and lattice structure. The present work clearly showed the dominant role of lattice disorder played over the paricle size effect in determining the magnetic dynamics of amorphous samples. We have also discussed some recent issues, such as: effects of magnetic clustering in the nature of magnetic phase transition, effect of spin-lattice interactions, unusual disorder magnetic states that are different from the conventional spin glass or superparamagnetism. The distinction between nanoparticle magnetism and amorphous magnetism, and properties related to structural and spin disorder, can be understood by extending similar experimental work to other systems. The concept of ferrimagnetism is applied to understand the core-shell magnetism in La$_{0.67}$Ca$_{0.33}$MnO$_{3}$ ferromagnetic nanoparticles.\\
\textbf{Acknowledgement:} We thank Pulak Roy for providing TEM data and CIF, Pondicherry University for providing SEM and EDAX measurements.

\begin{table*}
\caption{\label{tab:table2} The particle size of the milled samples are determined from the TEM data. \% Intensity is the normalized XRD peak intensity of 200 line compared to bulk sample. Structural notation: PCR = polycrystalline (bulk), NCR = Nanocrystalline, AMP= amorphous. Lattice parameters within $\pm$0.0003(a, b, c) are calculated from the XRD data.}
\begin{ruledtabular}
\begin{tabular}{cccccccc}
Sample & Milling hours & Size (nm) & \% Intensity & structural phase & a ($\AA$)
& b ($\AA$) & c ($\AA$) \\\hline
Bulk        & 0  & few $\mu$m & 100 & PCR & 5.4615 & 7.7203 & 5.4634 \\
mh25       & 25  & 65 & 87  & NCR & 5.4569 & 7.7323 & 5.4607 \\
mh61       & 61  & 12 & 14  & NCR & 5.4479 & 7.7576 & 5.4549 \\
mh98       & 98  & 16 & 9   & NCR + AMP & 5.4444 & 7.7749 & 5.4493 \\
mh146      & 146 & 60 & 5   & AMP       & - &- &- \\
mh200      & 200 & 90 & 3   & AMP       & - &- &- \\ 
\end{tabular}
\end{ruledtabular}
\end{table*}

\begin{table*}
\caption{\label{tab:table2} The fit parameters are obtained from simple Curie-Weiss law (equation 1) and from equation (3) for different samples. The effective paramagnetic moments $\mu$ and $\mu_{eff}$ are calculated from the Curie constant C and C$_{eff}$. The Curie temperature 
(T$_C$) and spontaneous magnetization at 0 K (M$_S$($\mu_B$)) are calculated from the Arrot plot analysis.}
\begin{ruledtabular}
\begin{tabular}{ccccccccc}
Sample & $\mu$ per f.u.($\mu_B$) & $\theta_w$ (K) & $\mu_{eff}$ per f.u.
($\mu_B$) & $\theta_1$ (K) & $\theta_2$(K) & $\xi$ (arb. unit) & T$_C$ (K) & M$_S$ (0)($\mu_B$) \\\hline
Bulk & 5.42 & 275 & -- & -- & -- & -- & 281 & 3.60 \\
mh25 & 5.75 & 240 & 7.20 & 150  & 252 & 91420 $\pm$875 & 262 & 2.17 \\
mh61 & 6.9  & 164 & 9.11 & -20 & 250 & 67200 $\pm$395 & 250 & 0.87 \\
mh98 & 5.43  & 120 & 6.84 & -54 & 248 & 90990$\pm$885 & 238 & 0.35 \\
mh146 & 5.33 & 55 & 6.20 & -59 & 245 & 64700$\pm$1200 & 225 & 0.17\\
mh200 & 5.33 & 30 & 6.11 & -70 & 244 & 53840$\pm$730 & 212 & 0.10\\ 
\end{tabular}
\end{ruledtabular}
\end{table*}
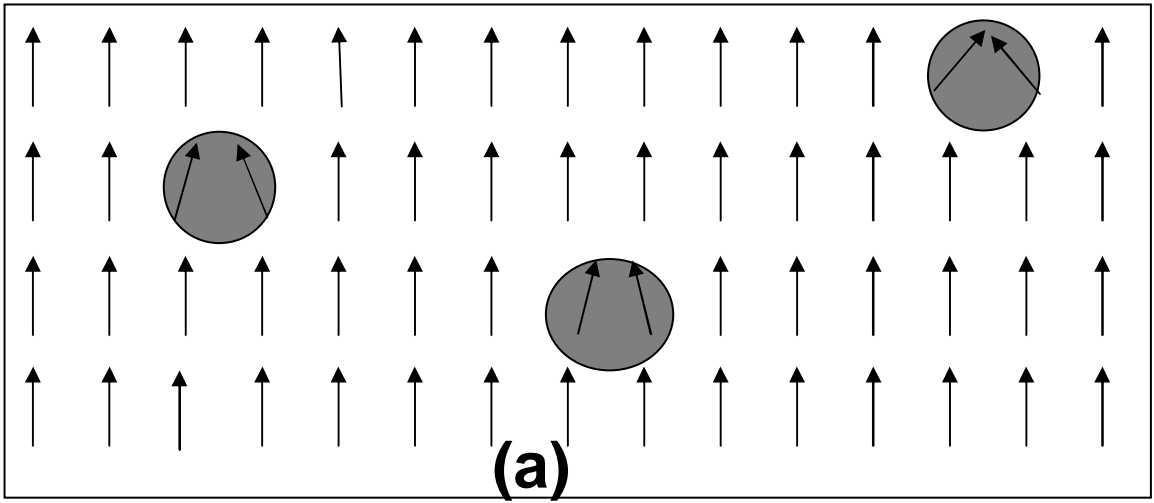
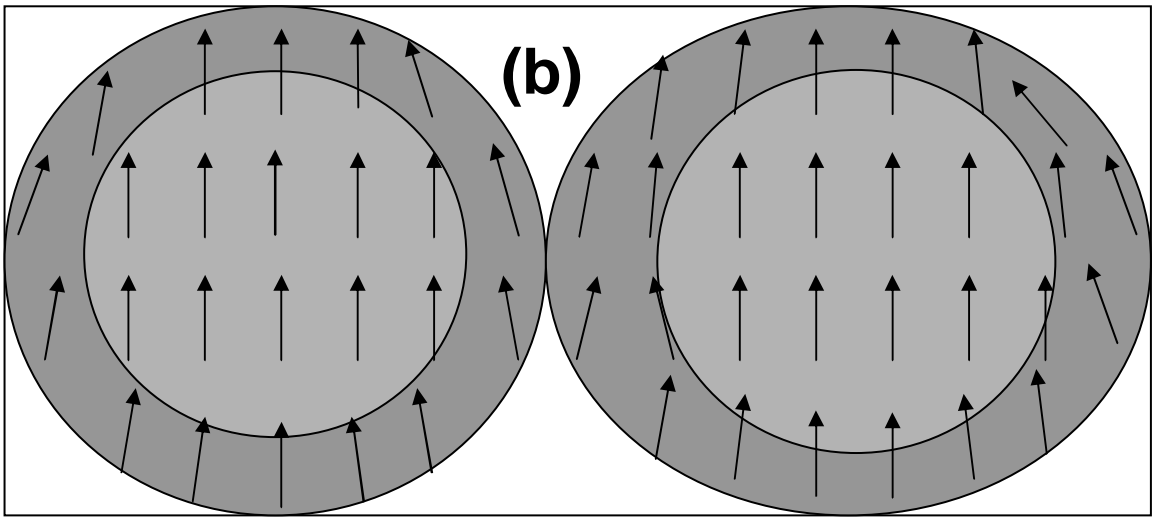
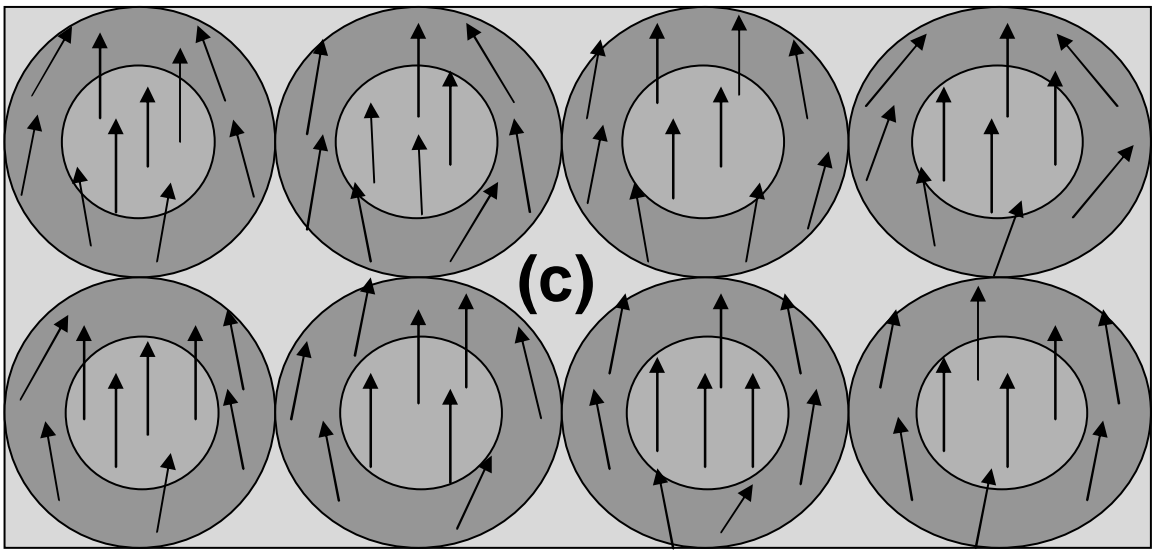
\end{document}